# Delineating Feminist Studies through bibliometric analysis

## Author information

Natsumi S. Shokida*, Diego Kozlowski** and Vincent Larivière***

*natsumi.solange.shokida@umontreal.ca
ORCID 0009-0008-9198-3888
University of Montreal, EBSI, Montreal, Canada
(corresponding author)

** diego.kozlowski@umontreal.ca
ORCID 0000-0002-5396-3471
University of Montreal, EBSI, Montreal, Canada

*** vincent.lariviere@umontreal.ca
ORCID 0000-0002-2733-0689
University of Montreal, EBSI, Montreal, Canada;
Stellenbosch University, Department of Science and Innovation - National Research Foundation Centre of Excellence in Scientometrics and Science, Stellenbosch, South Africa;
University of Quebec in Montreal, Observatoire des Sciences et des Technologies, Centre interuniversitaire de recherche sur la science et la technologie, Montreal, Canada

## Abstract

The multidisciplinary and socially anchored nature of Feminist Studies presents unique challenges for bibliometric analysis, as this research area transcends traditional disciplinary boundaries and reflects discussions from feminist and LGBTQIA+ social movements. This paper proposes a novel approach for identifying gender/sex related publications scattered across diverse scientific disciplines. Using the Dimensions database, we employ bibliometric techniques, natural language processing (NLP) and manual curation to compile a dataset of scientific publications that allows for the analysis of Gender Studies and its influence across different disciplines.

This is achieved through a methodology that combines a core of specialized journals with a comprehensive keyword search over titles. These keywords are obtained by applying Topic Modeling (BERTopic) to the corpus of titles and abstracts from the core. This methodological strategy, divided into two stages, reflects the dynamic interaction between Gender Studies and its dialogue with different disciplines. This hybrid system surpasses basic keyword search by mitigating potential biases introduced through manual keyword enumeration.

The resulting dataset comprises over 1.9 million scientific documents published between 1668 and 2023, spanning four languages. This dataset enables a characterization of Gender Studies in terms of addressed topics, citation and collaboration dynamics, and institutional and regional participation. By addressing the methodological challenges of studying "more-than-disciplinary" research areas, this approach could also be adapted to delineate other conversations where disciplinary boundaries are difficult to disentangle.

# Keywords

Feminist Studies, Gender Studies, Bibliometrics, Topic Modeling

# Introduction

During the 1970s, various lines of research previously scattered across different disciplinary communities became institutionalized and consolidated into Women's/Gender/Feminist Studies, characterized by its own scientific journals, departments, and training programs (Buker, 2003; Lykke, 2011; Richardson, 2010; Suárez Tomé, 2022). Gender Studies is often understood as an umbrella term for a diversity of inter-, intra-, inter-, post-disciplinary and even undisciplined approaches, and is often seen as a common political pursuit rather than a delimited object of study (Essed et al., 2005; Lundgren et al., 2015; Lykke, 2011; Prum, 2023). The field is also based on the heritage of Feminist Theory, Sexuality Studies, Gay and Lesbian Studies, Queer Studies/Theory, Trans Studies/Theory, Women's History, among others (Essed et al., 2005; Scott, 1986). Additionally, gender or feminist perspective may be adopted as an epistemological standpoint or theoretical framework for research within different disciplines, addressing power dynamics between genders in various domains (Gamba & Diz, 2021; Scott, 1986; Suárez Tomé, 2022). Given this "more-than-discipline" nature, feminist or gender/sex related studies present challenges for traditional bibliometric approaches.

Another peculiarity of this research area is that it is inherently linked to a social movement. Discussions around sex and gender, for example, are present both in the feminist movement and in its academic reflection in Gender Studies, but also permeate on fields like Biology and Medicine (Harding, 1991a). The demands of the women's and LGBTIQ movements for their representation and inclusion in clinical trials are also reflected in these disciplines (Epstein, 2009), and in epistemological debates around androcentric biases (Harding, 1991b; Maffia, 2007). In metascience studies, these discussions also appear as analyses of inequalities using gender/sex as an analytical category (Jacobs, 1996; Sugimoto & Larivière, 2023), explorations on the link between gender identities and research topics (Kozlowski et al., 2022; Picó-Pérez et al., 2024; Pradier et al., 2024; Sugimoto et al., 2019), and diagnosis and recommendations regarding the participation of women and LGBTQIA+ identities in science (Academy of Science of South Africa (ASSAf), 2016; Cislak et al., 2018; Schiebinger, 2000).

This means that feminist discussions prevalent in society are reflected in scientific literature, as gender/sex[1] related publications. Additionally, Feminist Studies can be understood as a particular case among different social movements that became disciplines, perspectives or currents of thought within existing disciplines, or broader conversations and research areas in science. Other examples include Black Studies, Human Rights, or Agroecology (Rojas, 2010; Suárez & Bromley, 2012; Wezel et al., 2009).

In this sense, Feminist Studies are a unique case of study to inquire the relation between science and society, and how a research area can have open discussions on epistemological definitions that transcend it. Nevertheless, this opportunity also comes with a challenge. Given that this thematic research transcends disciplinary boundaries, the identification of a relevant corpus cannot rely on established classifications found in bibliometric databases.

New computational tools, combined with manual curation, can contribute to identifying scientific publications related to Gender Studies even when they are scattered across multiple disciplines. In this paper, we apply bibliometric and natural language processing (NLP) techniques on the Dimensions database to construct a dataset of scientific publications that allows for the analysis of gender/sex related studies and its influence across different disciplines. Before presenting the methodology adopted and the results obtained, we will discuss the particularities of gender/sex related studies and reflect on the challenges they imply for their bibliometric operationalization.

---

[1] We use "gender/sex" (Fausto-Sterling, 2019) to refer to "the combined biological and cultural elements of human sex and social behavior" (Prum, 2023).

# Characteristics of Feminist Studies and challenges for its bibliometric study

While Gender Studies, Women's Studies, and Feminist Studies exhibit many characteristics of a discipline, various authors also emphasize their fluidity, which challenges traditional notions of disciplinarity. Their origins in the feminist movement, along with their social motivations and political commitments, underscore their unique position within academia. These studies do not have a monopoly over their central concepts, as "sex" and "gender" are analytical categories relevant to multiple scientific disciplines. In fact, scholars within Gender Studies acknowledge that part of their mission is to encourage the adoption of gender perspectives in other fields of knowledge (Mikkola, 2023), making these studies both a distinct area of inquiry and a bridge across various domains of knowledge. That is, feminist scholarship aims to impact different fields through the adoption of feminist epistemologies and critiques of biased research (Harding, 1991a). Feminist Studies are present in science as a whole, and intersect across diverse disciplines, including social sciences, humanities (Hassim & Walker, 1993; Scott, 1986), natural sciences (Longino, 1987; Richardson, 2010), medicine and primatology (Schiebinger, 2000). They address a wide array of problems, themes and concepts, including labor, patriarchy, gender identity, sexuality, kinship, modes of production, language, cultural representations and health, to name just a few (Epstein, 2009; Essed et al., 2005; Scott, 1986; Yun et al., 2020). This variability is due to the relationship between these studies with feminist movements, which have a broad agenda that does not necessarily coincide with the disciplinary boundaries of science (Hassim & Walker, 1993; Lykke, 2011). In addition to the thematic diversity, there is also an evolution in the meaning of central categories such as "gender" and "sex" (Scott, 1986).

Additionally, institutionalization appears as an important marker for defining a discipline, but the processes of institutionalization heavily depend on power dynamics, context, political situations, and the availability of resources. In countries and regions with fewer resources or political instability —such as Latin America (Ciriza, 2017) or South Africa (Hassim & Walker, 1993)—, there may be gender/sex related science without a strong institutionalization into Women's, Gender, or Feminist Studies.

These characteristics imply that feminist discussions within the scientific sphere exceeds the definition of a traditional scientific discipline. This can be represented as "overlapping circles" (Buker, 2003) between Gender Studies and other disciplines, as a "research area" that is more general than a discipline (Sugimoto & Weingart, 2015) and addresses topics of interest to the feminisms, or as "thematic research" and "thematic feminist studies" (Lykke, 2011). We will refer to this object as "gender/sex related studies" to emphasize their thematic focus on the category of gender/sex, or as "feminist studies" to highlight their connection with feminist activism and social movements. The thematic coherence of this research area acts as a "centripetal force" (Sugimoto & Weingart, 2015), positing Gender Studies as a discipline at the core of our inquiry.

Our goal is to bibliometrically delineate these gender/sex related studies. The difficulty in classifying this research area within conventional disciplinary boundaries highlights the need for approaches that do not rely solely on pre-established disciplinary classifications. In this case, it is more adequate to let the thematic research define itself "from the bottom up" (Sugimoto & Weingart, 2015).

Defining the scope of gender/sex related studies can be approached through several frameworks, including people-centered, publication-centered, and idea-centered methodologies (Sugimoto & Weingart, 2015). A people-centered approach would define the research area based on the community of researchers, emphasizing the contributions of authors and their institutional affiliations. Feminist Studies tend to be understood as a community (Buker, 2003; Lykke, 2011). However, the heterogeneous institutionalization of Women's/Gender/Feminist Studies in specific departments and programs, the dispersion of feminist research in different disciplines, and the impossibility of identifying the affiliation of researchers to feminisms, make the bibliometric approach focused on individuals less meaningful. A publication-centered approach, more common in bibliometrics, may categorize journals based on their titles and use citation patterns to map a field's structure (Tsay & Li, 2017), although this approach may impose a limiting structure on such a varied area. Feminism is expressed in science by having feminists as the subject of change in knowledge production, affecting the topics studied, the questions asked, and ways of answering them (Schiebinger, 2000). For delineating this research area, examining the contents of what is produced seems to be a more appropriate approach. An idea-centered approach would focus on keywords, specific language, thematic and methodological content of articles. Advances in NLP techniques make these approaches more popular, by revealing cognitive structures within disciplines and tracing networks of thematic influence (Milojević et al., 2011).

While various bibliometric studies have focused on gender in scientific production. Some of them are limited to analyzing the narrower field of Women's studies rather than Gender Studies and the gender perspective across various disciplines (Yun et al., 2020), or concentrate on identifying the most prestigious or productive journals in the discipline (Tsay & Li, 2017). Others are limited to one country (Söderlund & Madison, 2015) or a particular scientific domain (Dehdarirad et al., 2015; Lähdesmäki & Vlase, 2023). In other cases, they resort to using a limited list of keywords (Fox et al., 2022; Majumder et al., 2021). There is a need for a more comprehensive approach to bibliometrically delineate gender/sex related research. The historical evolution of terms like "sex" and "gender" implies significant changes in their meanings (Scott, 1986). Additionally, Lundgren et al. (2015) state that Gender Studies might be inspired or influenced by feminist theory without directly addressing gender, working on concerns that relate to gender sometimes centrally, sometimes peripherally. This makes it senseless to rely solely on terms like "sex", "gender" or "woman" to delineate gender/sex related studies. It is preferable to attempt to cover a broader spectrum of the issues concerning this field. Another central aspect when operationalizing a selection of Feminist Studies is coverage in terms of countries or regions and languages, as this research area is as international as feminist movements. In fact, for the analysis of agroecology (Wezel et al., 2009)—another scientific field related to a social movement—, taking into account different regional specific problems has led to interesting and rich answers. In some cases, "disciplines have multiple parallel 'births' in various countries" (Sugimoto & Weingart, 2015). If the intention is to have global coverage, then the limits will be given by the coverage of the database used. By using Web of Science (Dehdarirad et al., 2015; Fox et al., 2022; Lähdesmäki & Vlase, 2023; Tsay & Li, 2017; Yun et al., 2020), the analysis results in a strong bias towards North American and European countries. Maximizing coverage of other regions is important since the evolution of Gender Studies depends on local contexts (Rojas, 2010; Suárez & Bromley, 2012). For example, the cases of Latin America and Africa seem to present specificities in terms of the origin of these studies, their link with feminist activism (Vargas, 1993), and the intersection of gender issues with other dimensions such as race and class (Hassim & Walker, 1993). These specificities also appear in their possibilities of institutionalization, yielding different results in terms of research agenda and the characteristics of these scientific communities (Ciriza, 2017; Palacios-Nuñez, 2023). Basing the identification of Gender Studies on the prestige of journals or their productivity can also introduce unintentional omissions in this regard. Finally, if we pursue a corpus that is useful to analyze the historical evolution of Feminist Studies, it will be necessary to represent them from their beginnings to the present. This starting point seems to be agreed upon between the mid-60s and early 70s (Richardson, 2010; Scott, 1986; Suárez Tomé, 2022).

Given this dual nature of feminist science as a specific discipline—Women's/Gender/Feminist Studies—and as a perspective transversal across different disciplines, a bibliometric approach to achieve a representation of gender/sex related studies can benefit from a core-and-extension strategy. To delineate Feminist Studies, it makes sense to start with a core specialized in Gender Studies and then expand the search to other related publications. Since we consider it a discipline, the core can be identified through its journals. However, because the research area we want to delineate is thematically linked but broader than Gender Studies, it is appropriate to extract the topics covered in this core and identify them in other non-specialized journals. It is important to have temporal, regional and language coverage, as well as to meet quality standards for metadata. In the following section, we present the data source and methodological strategy.

# Data and methods

The data used for this paper comes from the Dimensions database (Herzog et al., 2020). This database was chosen because it has a broader coverage of literature from non-English speaking countries. Although this database does not completely index all scientific production, it is one of the best sources in terms of balance between a large coverage of journals (Visser et al., 2021) and metadata quality. In particular, it includes journals from non-English speaking countries and developing countries (Basson et al., 2022; Guerrero-Bote et al., 2021). In addition, indexing is not based on restrictive selection criteria (such as citations or reputation), but on the presence of a Digital Object Identifier (DOI). In Dimensions, classification by disciplines is available at the article level, and not at the journal level. We focus on publications in English, Spanish, French, and Portuguese, aligning with our language proficiency and capabilities.

The Dimensions database includes paper-level classifications at the division and group levels. Group "4405" corresponds to "Gender Studies". We analyzed the titles of a sample of documents classified under this group, and approximately only 63 out of every 100 documents align with what we defined here as gender/sex related studies. This lack of precision of the paper-level classification is also extended to the journal level. Considering the articles classified under this group, there are over 14,000 journals. We computed the proportion of articles classified under Gender Studies relative to the total number of articles published by the journals. Some of them are indeed specialized in Gender Studies but do not exhibit a high ratio, while others with a high ratio are not genuinely specialized in the discipline. For these reasons, we could not rely on this group for selecting relevant documents.

Given that gender/sex related studies encompass a conversation that extends across disciplines, we must include articles related even if they are not published in specialized journals. Following Lundgren et al. (2015), our methodological design starts with the definition of a core set of journals specialized in Gender Studies and then expanding it to articles from other journals. The expansion is made through topic modeling over titles and abstracts to retrieve the topics' salient terms, which are later used as a list of keywords for articles' retrieval. In other words, this approach places journals specialized in Gender Studies at the core, while the expansion also highlights the connection and impact of these discussions on the rest of the scientific output. This hybrid system surpasses basic keyword search by mitigating potential biases introduced through manual keyword enumeration. Simultaneously, the automatically generated topics are refined into an extensive keyword list through detailed manual review.

To select journals within the scope of Gender Studies across more than 85,000 journals indexed by Dimensions, we started with the list of journals categorized under "Women's Studies" in the Web of Science[2]. Secondly, we used indexes such as DOAJ[3] and LATINDEX[4] to ensure diversity and representation of journals beyond North America and Europe. Thirdly, we manually reviewed the journals that, according to Dimensions, had a majority of documents classified under the group "4405"-"Gender Studies". We completed the list by searching keywords within journal names. Recognizing that terms like "gender" or "sex" alone are insufficient to encompass the breadth of Gender Studies (Lundgren et al., 2015), an extensive keyword list was built manually and iteratively. Employing regular expression techniques, we identified journals whose names include the following terms across the four chosen languages: *gender, sex, woman, women, feminism, feminist, masculinities, lgbt, lesbian, gay, homosexual, bisexual, queer, girl*.

---

[2] Web of Science Group, Master Journal List https://mjl.clarivate.com/search-results, "Category: WOMEN'S STUDIES", last accessed March 30th, 2022.
[3] DOAJ (Directory of Open Access Journals) https://doaj.org/search/journals, "Subject: Women. Feminism", last accessed March 30th, 2022.
[4] LATINDEX (Regional Online Information System for Scientific Journals of Latin America, the Caribbean, Spain and Portugal) https://www.latindex.org/, "Advanced search: Directorio, Subtema: Estudios de género", last accessed March 30th, 2022.

This list of journals was then manually curated to ensure they belong to what is traditionally considered as Gender Studies. For this core set, we focused on Social Science and Humanities' journals. The journals belonging exclusively to Medical and Health Sciences, Biological Sciences, Engineering, and Psychology and Cognitive Sciences were excluded from this stage, and will eventually reappear in the extended set of documents. Nevertheless, the filter by discipline was not categorical. For example, some journals with focus on LGBTIQ+ population were partially associated with Psychology and partially with Humanities, and were kept as part of the core. On the contrary, exclusively medical journals focused on females' bodies (obstetrics, gynecology, radiology, etc.) were not included in the core. As a result, the core corpus comprises 289 scientific journals.

From the 289 journals selected as the core of our dataset we extracted 160,030 articles spanning from 1970 to 2023. Among these, more than 135,000 articles have a title, and more than 65,000 include abstracts. The extended corpus is based on keyword search, comprising all journals in Dimensions. The list of keywords for this retrieval was crafted using topic modeling on the core set. Topic modeling is a non-supervised technique that helps to identify the subjects of a corpus, their distribution, and their most salient terms. In this case, we use these models as a tool for the construction of a keyword list because the topics' salient terms function as a very detailed list of specific vocabulary.

Applying BERTopic (Grootendorst, 2022) to this corpus, we generated a list of 330 topics with at least 50 publications in them, along with their characteristic words. In Table 1, we provide a few examples of these topics. The model generates topics that blend terms from various languages while maintaining thematic coherence. For instance, in topic #31, the words "prostitution", "trafficking" and "human" are in English, whereas "sexualidad" represents "sexuality" in Spanish.

**Table 1.** Topics with their specific words, and number of documents.

| Topic | Count |
|---|---:|
| 0_queer_lesbian_gay_lgbt | 2,091 |
| 27_masculinity_masculinities_man_masculine | 263 |
| 31_trafficking_human_prostitution_sexualidad | 248 |
| 82_pregnancy_maternal_motherhood_mother | 139 |
| 93_feminismo_feministas_feminismos_brasil | 133 |
| 94_black_race_justice_freedom | 133 |
| 138_labor_workers_migrant_care | 101 |

Each topic is associated as a set of keywords, and we will use them to create a comprehensive list of keywords that are relevant to Gender Studies. Nevertheless, not all keywords are useful for this task. Therefore, a meticulous manual analysis of the topics was necessary to identify and compile these keywords. The resulting keyword list is a combination of terms that were already used for the journal search (e.g., "feminist", "woman", "gender"), new terms that emerged from the topic modeling (e.g., "homophobia", "menstruation", "abortion"), and reconfiguration of terms that appeared in the topics but were too generic as standalones. For example, words such as "violence" were reconfigured to more specific expressions such as "gender based violence" or "obstetric violence".

This combined process of automated term detection and manual curation proves highly effective, as it reveals keywords that were not initially considered during the core-journals identification phase. For instance, the absence of the term "heterosexual" initially, in contrast to terms like "homosexual", "lesbian", or "bisexual", highlights a researcher bias towards what is conventionally seen as default versus specific. This also shows how the methodology devised serves as a tool to address and correct such biases.

Furthermore, incorporating a manual review stage allows for the discovery of new terms that were neither initially anticipated nor generated by the model. Terms such as "misogyny", "travesti", or "transphobia" are included based on the appearance of related terms, by means of the domain knowledge of the research team. To ensure comprehensive coverage across the four chosen languages in the keyword search across the entire database, the list of terms is meticulously expanded. This process culminates in a final list of 229 keywords[5], which were used to conduct keyword searches within article titles, disregarding their abstracts. This search yielded a total of 1,894,425 documents from 1668 to 2023, of which 1,807,272 are new additions to the core.

Overall, the dataset comprises 1,967,302 documents. Among these, 160,030 articles (8.1%) originate from journals specifically dedicated to Gender Studies topics (the core), while 1,807,272 articles (91.9%) address topics related to those discussed in the core, identified through keyword searches of their titles.

# Results

The corpus comprises over 1.9 million scientific documents published between 1668 and 2023, sourced from the core set or documents featuring one or more of our selection of keywords in their titles. While the majority of those are research articles, the corpus is not limited to this document type. The corpus is segmented according to the following distinction: 1) the Core includes articles from Gender Studies journals, primarily those labeled by Dimensions as from Social Sciences and Humanities; 2) the rest (Not Core), comprises articles identified through keyword searches, excluding those overlapped with the Core.

Table 2 provides absolute and relative sizes of these segments. As expected, the Core features fewer journals but more articles per journal. The Not Core segment is much larger, containing mostly articles from Biomedical and Clinical Sciences, Health Sciences, and Biological Sciences. This result reflects the higher publication rates in these disciplines compared to Social Sciences and Humanities, as well as a higher average number of citations (Sugimoto & Larivière, 2018).

**Table 2.** Composition of the dataset.

| Segment | Articles (A) | Journals (J) | Articles per journal (A/J) | Average citations |
|---|---|---|---|---|
| Core | 160,030 (8.1%) | 282 (0.5%) | 567 | 8.8 |
| Not Core | 1,807,272 (91.9%) | 60,637 (99.5%) | 30 | 13.1 |

---

[5] Taking into account the variations incorporated to perform the query, there are a total of 259 different regular expressions. For simplicity and interpretability, for the analysis of results these keywords were organized into 63 grouping terms in English.

| | Total | 1,967,302 | 60,919 | 32 | 12.8 |
|---|---|---|---|---|---|

Expanding the core specialized in Gender Studies to include publications from other journals resulted in a much larger corpus, yet still connected to the topics addressed in the core. As shown in Figure 1, more than half of the Core overlaps with the documents identified through keyword retrieval. This shows that the keywords used were indeed characteristic of the discipline.

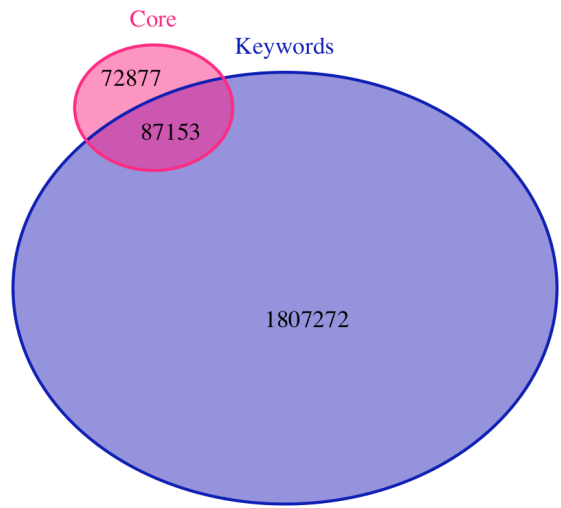

**Figure 1.** Overlapping between the core and the rest of the dataset.

The earliest publications in journals specifically focused on Gender Studies that appear in Dimensions date back to 1970. The publication volume in the Core experienced a rapid growth until 1975, after which it stabilized at a growth rate similar to the rest of the dataset. This trend aligns with literature on the institutionalization of Feminist Studies after the 'second wave' of US feminism. The era of 'radical feminism' during the 1960s and 1970s marked a prolific stage primarily characterized by books authored by figures like Kate Millet, Betty Friedan, or Simone de Beauvoir from Europe. Concurrently, the 1960s witnessed the inception of Feminist Science Studies as "a multidisciplinary stream of feminist scholarship on gender and science" (Richardson, 2010). In the 1970s, Feminist Epistemology emerged as a philosophical branch (Richardson, 2010; Suárez Tomé & Maffía, 2021), alongside materialist feminist movements in Latin America (Suárez Tomé, 2022). It is after this stage that critical disciplines —such as Sexualities Studies, Gay and Lesbian Studies, Queer Studies, Trans Studies— sprout, to later contribute to the broad arc of Gender Studies. The significant increase in the core set around the mid-1970s reflects the institutionalization and consolidation of research lines previously dispersed across various disciplinary communities (Lykke, 2011; Richardson, 2010; Scott, 1986). This is confirmed by the fact that there were already publications related to these topics well before this period. Figure 2 illustrates the evolution of the number of articles within each segment: from 1976 to 2022 for the Core, and from 1950 to 2022 for the Not Core. In this case, the periods for each segment were defined based on the stabilization of the annual number of publications. It is important to note that the increase in scientific production during these decades is not unique to Gender Studies but is part of a broader trend (Sugimoto & Larivière, 2018).

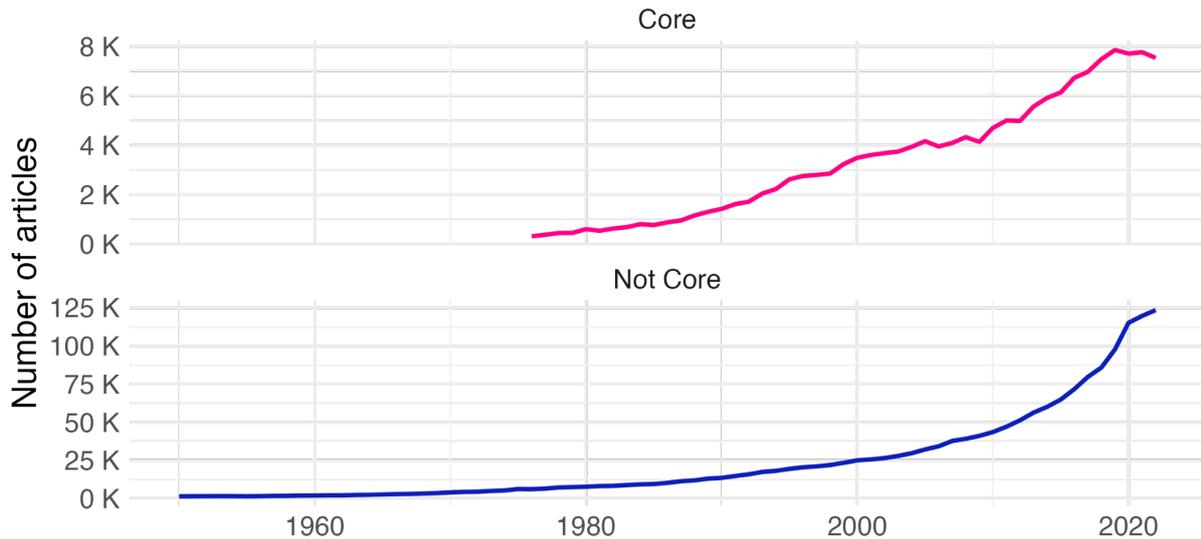

**Figure 2.** Evolution of the number of articles in each part of the dataset
(Core: 1976-2022, Not Core: 1950-2022)

Regarding the disciplines associated with each article (Figure 3), it is noteworthy that the Core features a significant proportion from Human Society (69%) and History, Heritage, and Archaeology (6.8%). Conversely, the Not Core segment is more focused on Biomedical and Clinical Sciences (37.2%), Health Sciences (12%), and Biological Sciences (6.1%). Nevertheless, it also includes documents from Human Society (13.6%), while Psychology represents between 5% and 7.2% in both segments. Given the relative size of the Not Core segment within the overall dataset, the general distribution closely resembles that of this segment. This asymmetrical distribution of disciplines across each segment of the dataset should be considered when analyzing the distribution of keywords.

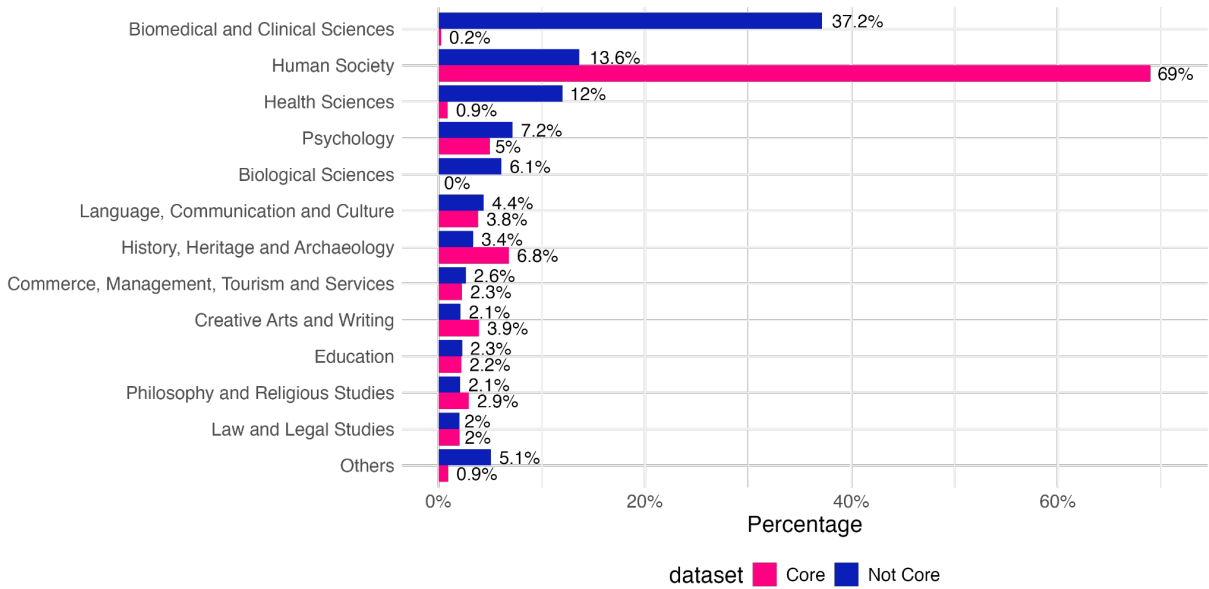

**Figure 3.** Distribution of disciplines in each part of the dataset
*(<2% for both were grouped in "Others").*

Figure 4A illustrates the most frequently used keywords in each segment of the dataset. The keywords used for the queries were consolidated into 63 representative terms. That is, keywords referring to the same topic, expressed in different languages, were grouped under a single term. Notably, in both segments, the top three terms consistently identified are "women", "sex", and "gender". Interestingly, "gender" appears as a more prevalent keyword in the Core, while "sex" is more prominent in the remainder of the dataset. Furthermore, the term "feminis"[6] is prominently ranked within the Core but is less frequently encountered outside it. This divergence reflects the influence of Feminist Theory, emphasizing a political commitment and focusing on discussions surrounding gender roles in society, which are more pronounced in Gender Studies and the broader fields of Social Sciences and Humanities compared to Medicine or Biology.

The temporal evolution of these keywords is depicted in Figure 4B. By examining the relative frequency, which represents the number of documents containing a keyword relative to the total number of papers in each segment, we can see that the higher prevalence of "women" and "woman" in the Not Core, compared to the Core, remains steady over time. Additionally, the increasing attention to masculinities (especially in the Core during recent decades) and gender violence is notable. The term "queer", associated with the contributions of Queer Theory, Sexuality Studies, and the LGBTIQA+ community, shows an upward trend since the 1990s, particularly in the Core. While the use of terms related to homosexuality decreases, more inclusive acronyms such as 'LGBT' gained popularity in the 20th century. Terms like "lesbian" and "gay" are more frequent in the Core (Figure 4A), though they exhibit fluctuations over time (Figure 4B). Terms related to trans people rank among the top 10 most frequent terms, reflecting both medical and health discussions concerning the trans population (Richardson, 2022), as well as the significant relevance within the core of Gender Studies.

Terms such as "abortion" and "rape" are more prevalent outside the Core. These may refer to issues that constitute a central part of the feminist agenda globally, but on which the Gender Studies community might already have reached some consensus, so we do not capture publications related to this topic as a prevalent ongoing debate at the core. Nevertheless, they may remain as important topics outside the core, for its practical social and medical implications. The term "menstrual" is also more prevalent outside the Core, which is expected due to its relevance for Health Sciences.

---

[6] "feminis" captures words such as "feminism", "feminist", and their variations in the 4 selected languages.

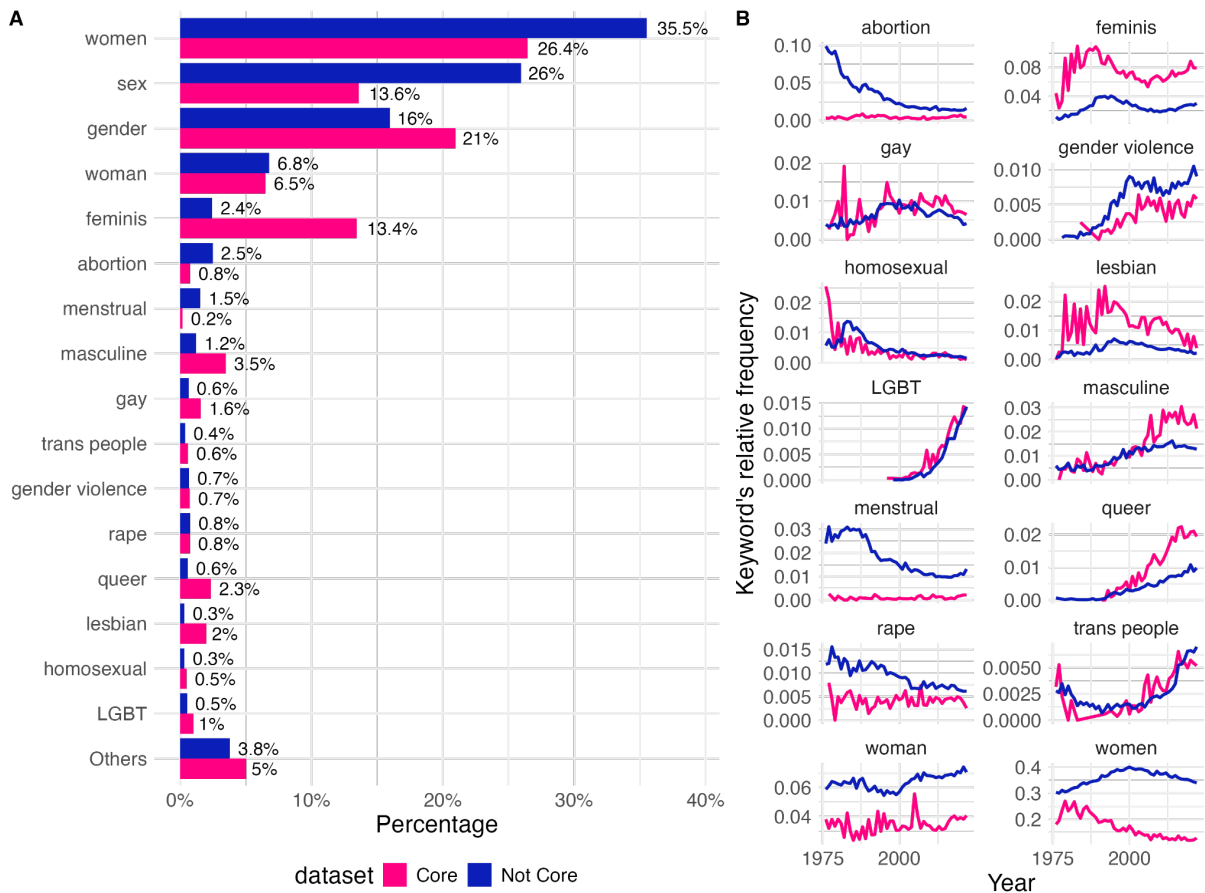

**Figure 4.** A: Relative frequency of keywords in each part of the dataset; B: Evolution of the relative frequency of a selection of keywords.

Figure 5A compares the evolution of the relative frequencies of the terms "sex" and "gender" across each segment of the dataset. The popularization of the term "gender" is evident, particularly during the 1980s. In the Core segment, "gender" surpasses the use of "sex". In the rest of the dataset, which is largely composed of works from Health and Biomedical and Clinical Sciences, the curves do not intersect, though they come close to converging. As Scott (1986) points out, these terms evolve in their meaning and use. The popularization of the term "gender" and the decline in the use of "sex" was interpreted by Haig (2004) as a replacement. However, Figure 5B shows a modest but steady growth in the concurrent use of both terms in the titles of publications. This probably reflects the discussions on biomedicine on how "sex and gender interact and co-constitute each other along multiple dimensions" (Richardson, 2022), the abandonment of single-axis frameworks of sex and gender, and the adoption of the intersectional concept of gender/sex (Fausto-Sterling, 2019; Prum, 2023).

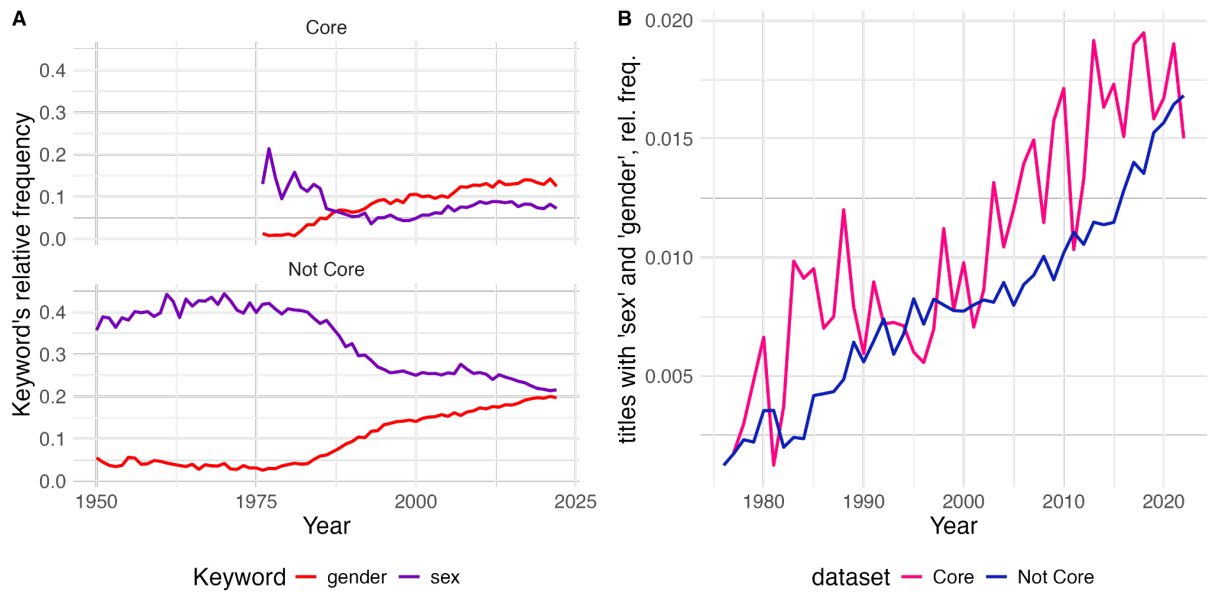

**Figure 5.** A: Keywords "sex" and "gender", evolution in each part of the dataset; B: Evolution of the usage of sex and gender in titles, simultaneously.

# Discussion

This paper presents a methodology for creating a bibliometric corpus on gender/sex related studies. As a result, we built a dataset of over 1.9 million articles, in four different languages (English, Spanish, French, and Portuguese), published between 1668 and 2023. The corpus is divided into core Gender Studies publications, and articles from different disciplines. This is accomplished by combining bibliometrics and NLP techniques, with manual curation. These techniques complement each other to avoid biases or gaps, both in the model's outputs and during the reflexive instances. Furthermore, the methodological strategy divided into two stages (first focusing on a specialized core and then extending to the rest of the publications), reflects the dynamic interaction between Gender Studies and its dialogue with different disciplines.

In general, the limitations encountered when working with bibliometric databases are primarily related to the data source and its characteristics (in this case, Dimensions). This includes issues such as the heterogeneous coverage across disciplines, languages, and regions, as well as the limited inclusion of literature beyond journal articles (Archambault & Larivière, 2010; Sugimoto & Larivière, 2018). In our case, concerning regional coverage, the comprehensive inclusion of Latin American scientific journals dedicated to Gender Studies allows for a good representation of this region in the core dataset. However, the same cannot be said for regions like Africa or Asia. Additionally, it's important to note that our analysis does not encompass the complete literature of Gender Studies, given the central role of books, essays and pamphlets in Feminist Theory (Suárez Tomé, 2022). A possible extension of this project could delve into the reference lists of articles to catalog those documents.

Gender Studies and gender/sex related studies can be seen as a specific case among different social movements that have entered academia —in this case, the feminist agenda—, impacting scientific discussions across diverse disciplines (Rojas, 2010; Suárez & Bromley, 2012). These discussions carry significant social relevance, and it is our goal to offer a characterization of their contributions to society as a whole. The methodology presented will enable a deeper study of a knowledge area that is challenging to address within the traditional disciplinary structures found in bibliometric databases. This approach could also be adapted to delineate other conversations where disciplinary boundaries are difficult to disentangle.

The dataset obtained from this work allows advances in the characterization of feminist research in terms of citation and collaboration dynamics, the involvement of institutions, countries, and regions, as well as the emergence, popularization, and decline of certain concepts or theories. Our future lines of research include large-scale topical and network analysis. In this way, the conversation on the role of gender/sex in society would have a map of itself, allowing us to observe which subjects are more or less explored, by whom, in which regions or eras, or even in which languages.

# Declarations

## Data transparency

Once the article is accepted, the following listings will be published in a public repository: list of 289 scientific journals comprising the core set of Gender Studies; the list of 300 topics obtained as a result of the topic modeling; the list of 259 keywords used for data-retrieval; the list of the DOIs of the articles that compose the obtained dataset.

## Author contributions

Conceptualization, Writing – review & editing: Natsumi S. Shokida, Diego Kozlowski, Vincent Larivière
Data curation, Project administration, Validation, Writing – original draft: Natsumi S. Shokida
Formal Analysis, Methodology, Software, Visualization: Natsumi S. Shokida, Diego Kozlowski
Funding acquisition, Supervision: Vincent Larivière

## Competing Interests

The authors have no competing interests to declare.

## Funding

Diego Kozlowski and Vincent Larivière acknowledge funding from the Social Science and Humanities Research Council of Canada Pan-Canadian Knowledge Access Initiative Grant (Grant 1007-2023-0001), and the Fonds de recherche du Québec - Société et Culture through the Programme d'appui aux Chaires UNESCO (Grant 338828). Natsumi Shokida acknowledges funding from the UNESCO Chair on Open Science.